\documentclass[12pt,a4paper,twoside]{scrartcl}
\usepackage{graphicx}
\usepackage{amsmath}
\usepackage[figuresright]{rotating}
\usepackage{fancyhdr}
\usepackage[width=16cm]{geometry} 
\def\skiplinehalf{\medskip\\}

\title{Selected Aspects of Neutron Decay}
\author{Stephan Paul
\skiplinehalf Physik Department E18\\
Technische Universit\"at M\"unchen \\
James Frank Strasse, D-85748 Garching}
\date{}

\newcommand{\ket}[1]{\vert#1\rangle}
\begin{document}
\maketitle
\begin{abstract}
Precision measurements of neutron decay offer complementary access
to particle physics at small distance scales or high energies. In
particular they allow tests of the V-A structure of the weak
interaction. Among many experimental activities which are ongoing
around the world we present two new experiments which are planned or
studied for the near future. While the neutron lifetime still bears
significant experimental uncertainties and thus has to be studied
with greatest precision the two-body decay ($n\rightarrow
H\overline\nu$) of the neutron has not yet been observed. Despite
its small branching fraction it offers many possibilities in the
framework of particle physics. Both cases are addressed in this
contribution.
\end{abstract}%

\section*{Introduction}

The process underlying neutron decay is the semileptonic transition $%
d\rightarrow ue^{-}\overline{\nu }$. The Hamiltonian for this process is
written as

\begin{center}
$H=\ \frac{G_{F}}{\sqrt{2}}V_{ud}\cdot \overline{e}\left( 1-\gamma
_{5}\right) \nu \cdot \overline{u}\left( 1-\gamma _{5}\right) d$ \ with $%
\frac{G_{F}}{\sqrt{2}}=\frac{g^{2}}{8M_{W}^{2}}$ \ and $g\cdot \sin
\vartheta _{w}=e\label{Hamiltonian_sl_decay}$
\end{center}

and describes the usual V-A coupling known for weak interaction. Here, $%
V_{ud}$ \ describes the quark mixing (see CKM\ matrix), $G_{F}$ is
the Fermi coupling constant, $M_{W}$ is the mass of the W-boson,
$\sin \vartheta _{W}$ the Weinberg angle and e the electric charge.
The two terms describe the lepton and quark transition amplitudes.
However, as there is no free quark decay the process to be
considered is more complicated (see fig.1
).

\begin{figure}[ht]\begin{center}
\includegraphics[
height=1.7417in, width=2.1854in]
{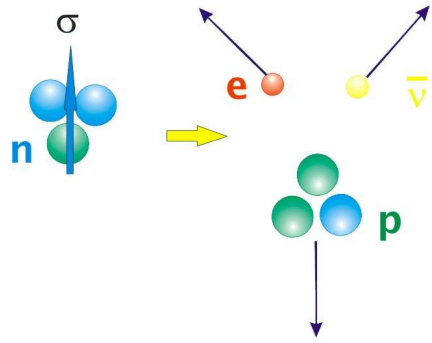}\caption{%
Sketch of the standard neutron decay}\label{n_decay}%
\end{center}\end{figure}%

In the hadronic environment the coupling constants turn into
formfactors, which however are evaluated at very small
$q^{2}\rightarrow 0$. The quark part of the Hamiltonian thus has to
be rewritten as

\vspace{.5cm}

\begin{center}
$V_{\mu }+A_{\mu }=V_{ud}\cdot \overline{\psi _{p}}\left( f_{1}\left(
q^{2}\rightarrow 0\right) \gamma _{\mu }+f_{2}\left( q^{2}\rightarrow
0\right) \frac{\sigma _{\mu \nu }}{m_{p}}q^{\nu }+f_{3}\left(
q^{2}\rightarrow 0\right) \frac{q^{\mu }}{m_{p}}\right) \psi _{n}+\overline{%
\psi _{p}}\left( f_{i}\rightarrow g_{i}\right) \psi _{n}$
\end{center}

\vspace{.5cm}

Using conservation of vector current and evaluating the expression at $%
q^{2}\rightarrow 0$ leaves only two terms

\begin{itemize}
\item $G_{V}=f_{1}\left( q^{2}\rightarrow 0\right) \cdot G_{F}\cdot
V_{ud}=g_{V}\cdot G_{F}\cdot V_{ud}${}

\item $G_{A}=g_{1}\left( q^{2}\rightarrow 0\right) \cdot G_{F}\cdot
V_{ud}=g_{A}\cdot G_{F}\cdot V_{ud}$
\end{itemize}

\noindent all other contributions vanish. For simplicity we define
another quantity, called $\lambda $, with

\begin{center}
$\lambda =\frac{G_{A}}{G_{V}}=\frac{g_{A}}{g_{V}}$
\end{center}

If we take $G_{F}$ from muon decay we are left with two independent
'free' parameters, $V_{ud}$ and $\lambda$. These parameters can be
determined experimentally combining different measurements in
neutron decay, decay asymmetries and the total decay rate using the
neutron lifetime. The latter one relates to our free parameters via

\vspace{.5cm}

\begin{center}
$\tau \propto \frac{1}{G_{V}^{2}\cdot \left( 1-3\lambda ^{2}\right) }$
\end{center}

\vspace{.5cm}

\noindent Experimentally two different approaches have been taken to
measure the neutron lifetime.

\begin{itemize}
\item Until the late eighties most experiments have used the\textit{\ in-beam%
} method which requires the detection of n-decay products from a
well defined fiducial volume. Although being competitive on the
statistical accuracy this method faces many systematic problems
which has limited the final precision of the experiments.

\item The most precise experiments have been performed using the method of
\textit{stored neutrons}. Ultra cold neutrons (UCN) can be confined
in bottles by means of their interaction with the surface, the
gravitational field and magnetic field gradients. Choosing proper
materials neutrons are reflected from the surface with minimal
losses which may occur either by absorption or by up-scattering (in
which case they gain sufficient energy to leave the containing
volume). The maximal allowed energies of such neutrons is around
250-300 neV. The neutron lifetime is derived from the number of
neutrons extracted from this bottle after various storage times t.
Varying the momentum spectrum of the neutrons one can estimate the
loss rate in the bottle due to effects other than weak decay. These
experiments have the virtue of high statistical accuracy (though
still limited by the strength of the UCN sources available). The
average lifetime extracted from many of such experiments is
$885.7\pm 0.8s$ \cite{PDG_2004}. Very recently, however, a new
result has been published deviating from the present world average
by about six standard deviations \cite{Serebrov_2004zf}. As the
basic setup of this last experiment is almost identical to one of
the previous measurements \cite{Nesvizhevsky_1992ej} it indicates
that systematic effects, although claimed by all authors to be well
under control, still are a major issue.
\end{itemize}

\begin{figure}[ht]\begin{center}
\includegraphics[
height=3.1349in, width=5.0886in]
{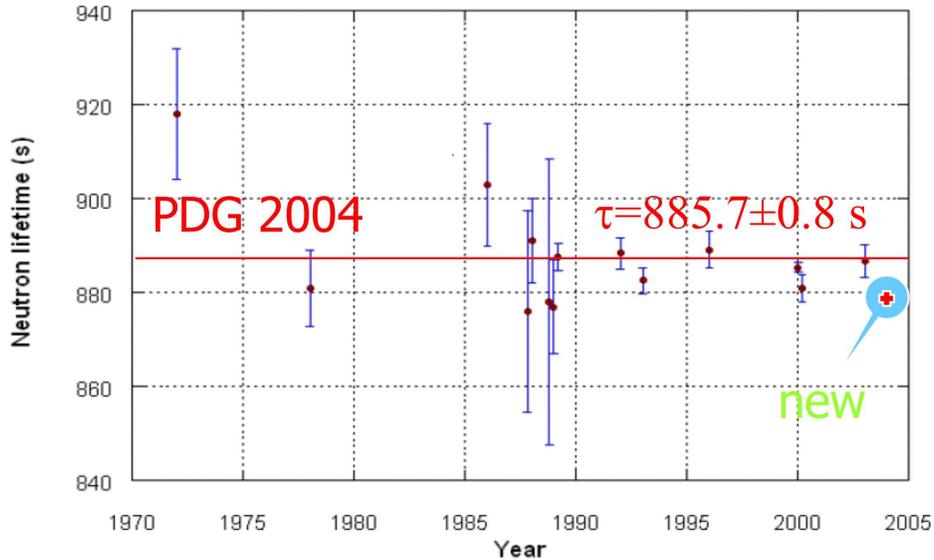}\caption{%
Experimental results for the neutron lifetime and the variation with time.
Note the latest result which deviates from the world average by about 6
standard deviations.}\label{n_lifetime}%
\end{center}\end{figure}%

We can now combine the results for the n-lifetime experiment as well
as the neutron decay asymmetries for the electrons (not discussed
here) to extract the quantities $V_{ud}$ and $\lambda $. This is
shown in fig. 3, taken from \cite{Abele_2003ya}. $V_{ud}$ can also
be determined by other methods, e.g. superallowed $0^{+}\rightarrow
0^{+}$ nuclear beta decays, which offer superb statistical accuracy
but suffer from systematic uncertainties connected to corrections
for the nuclear medium. In addition, we can compare
the direct measurements for $V_{ud}$ with the results from $V_{ub}$ and $%
V_{us}$ in combination with the unitarity of the CKM matrix. Both
values are also shown in fig 3. The present data set shows a clear
disagreement of the different methods. This can have several
reasons. It may hint to a non-unitarity of the CKM matrix or to the
imcompleteness of the Hamiltonian. Thus, other contributions like
right-handed currents or scalar and tensor interactions could play a
role. On the other hand, experimental problems are not yet excluded
as can be seen from the recent lifetime results which by itself
would cure the present disagreement. Very recently a new analysis of
neutral K-decays (K$_{l3}$) has given new values of $V_{us}$, higher
than the previous results which in turn lowers the expectations for
$V_{ud}$ values extracted from the unitarity assumption
\cite{Mescia_2004xd}. But also there, systematic uncertainties,
mainly related to theoretical calculations in the framework of
chiral perturbation theory are at play. Taken at face value, the
most recent results in kaon physics would by itself also cure the
disagreement but would disfavor the new lifetime result.

\begin{figure}[ht]\begin{center}
\includegraphics[
height=3.1592in, width=4.0456in]
{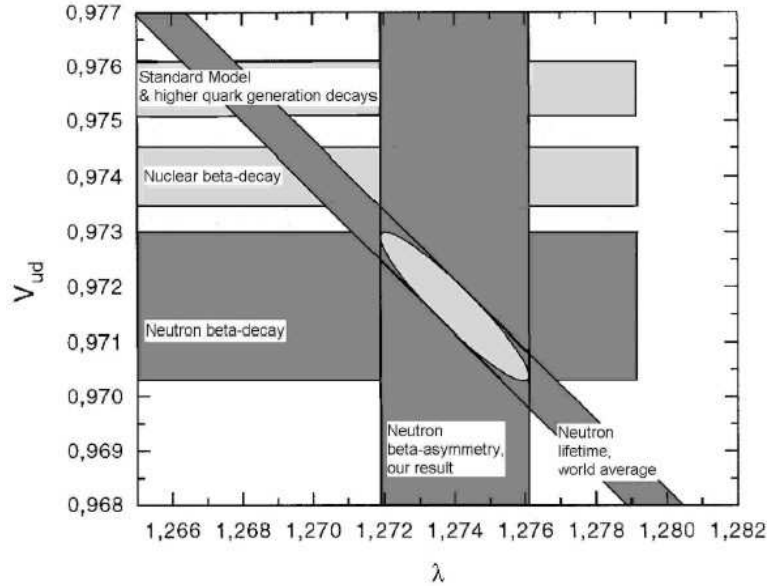}\caption{%
Experimental determination of the two standard parameters in neutron decay, V%
$_{ud}$ and $\protect\lambda $ $\protect\cite{Abele_2003ya}$ .}
\label{standard_model_parameters}%
\end{center}\end{figure}%

It becomes clear from this discussion that new measurements are needed with
emphasis on the understanding of systematic effects.

\section{A New Neutron Lifetime Experiment}

A few years ago the group of the TUM has proposed a new experiment
to measure the n-lifetime based on the use of a magnetic trap for
ultra cold neutrons \cite{Hartmann}, \cite{Picker}.This trap offers
the unique possibility to simultaneously detect decay protons in
real time as well as neutrons which had suffered spin flip during
storage, the only foreseeable loss mechanism in a magnetic trap. The
principle of this trap is shown in fig. 4. The trap consists of 19
super-conducting coils powered with alternating current direction
thus forming a magnetic multipole system with inner diameter of 50
cm. The forces in such a trap acting on the neutron are described by

\begin{center}
$\overrightarrow{F}=-\overrightarrow{\nabla }\left( \overrightarrow{\mu }%
\cdot \overrightarrow{B}\right) $
\end{center}

\noindent where the magnetic moment of the neutron $\mu =-60.5 \rm
neV/T$ results in a repulsive force when $\overrightarrow{\mu }$ and
$\overrightarrow{B}$ are parallel. Neutrons with this spin
orientation are repelled from the walls which are at a maximal
magnetic field of about 2 T. In order to avoid spin flip processes
in low field regions a cylindrical field is superimposed which is
generated by a current rod in the center of the cylindrical trap.
The current rod is shielded from the neutrons again by means of a
multipole field generated by 18 current loops. The upper lid of the
trap, which is at a height of about 1.2 m, is formed by the
gravitational force. Neutrons with
energies up to 120 neV can be stored in the trap.%

\begin{figure}[ht]\begin{center}
\includegraphics[
height=4.0352in,
width=2.0116in]
{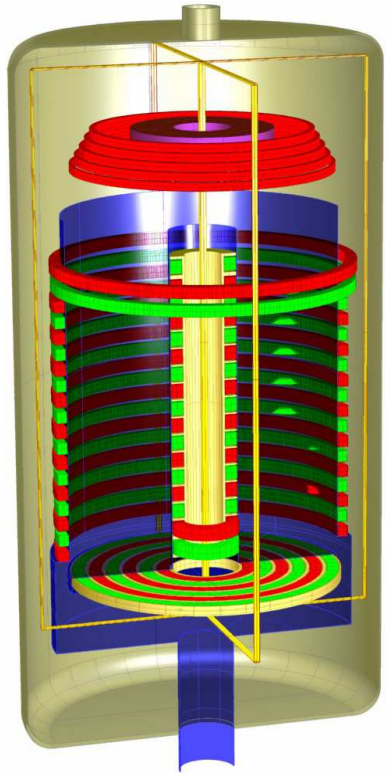}%
\caption{The proposed neutron lifetime experiment of the TUM.
Neutrons can be stored inside a magnetic trap formed by two
concentric cylindrical multipole fields. The trap can be filled
(emptied) by a slit at the bottom which is open for all neutrons
when the magnetic field is ramped down.}
\label{lifetime_experiment_1}%
\end{center}\end{figure}%


\qquad Neutrons are filled into the trap from a slit in the lower
part of the setup which is connected to the UCN buffer volume (UCN
source). It is situated between the outer torus and the bottom lid.
The slit can be opened and closed magnetically by ramping the field
structure within 100 s.

In the first 150s of the filling neutrons are stored by means of a
reflective layer mounted on the substrate which also serves a
generator for an electric field. A thin cylinder surrounds the inner
and outer coils separating them mechanically from the storage
volume. On the side pointing away from the storage volume a
conductive structure is mounted to form the field lines for the
electrical field used to collect the decay protons.The inner side
will be coated with a neutron reflective material. The initial phase
of the storage cycle is also used for spectrum cleaning. Fig. 5
(right) shows the typical path of a neutron in the bottle when
operated in storage mode. The path of decay protons is shown in fig.
5 (left). They are extracted towards the top by means of an
electrical field of 20 kV and a magnetic field with a component
along the electrical field lines. After passing a focussing magnet
(at the top) they are further accelerated onto a proton detector.
\begin{figure}[ht]\begin{center}
\includegraphics[
height=4.203in, width=6.2491in]
{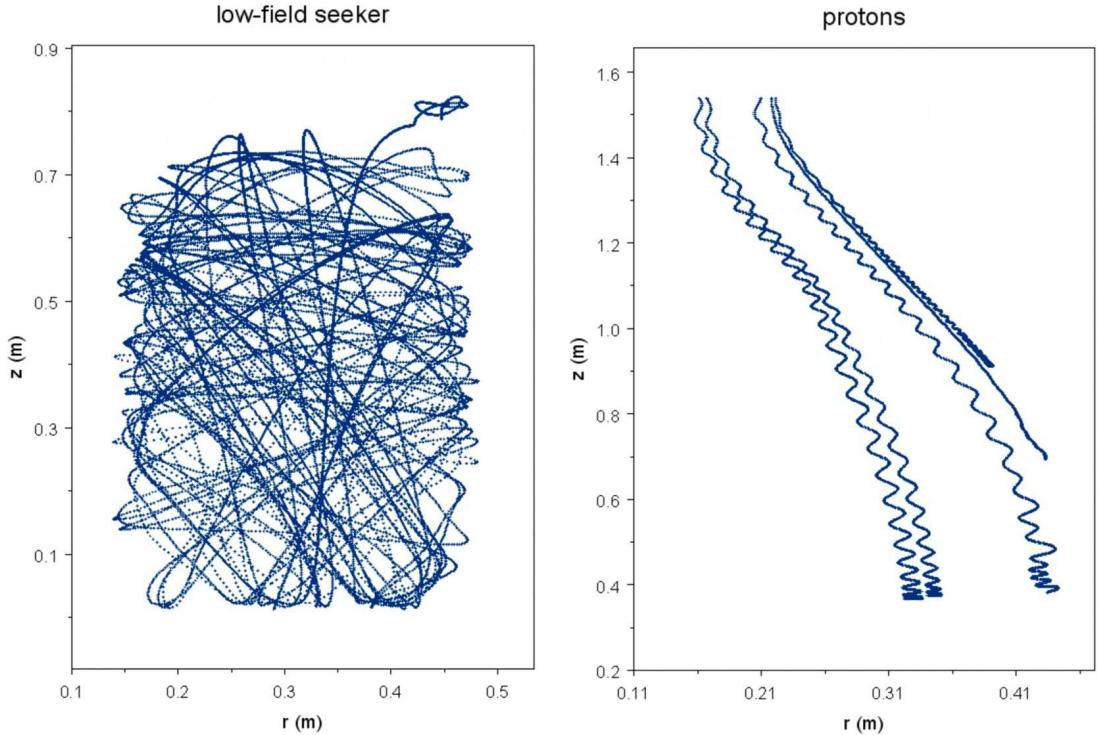}%
\caption{Left: Result of a ray tracing program for a neutron stored
inside the magnetic trap. Right: Result of a ray tracing program for
decay protons inside the
magnetic trap.}\label{particle_traces}%
\end{center}\end{figure}%

The trap is foreseen to be installed at one of the newly planned
high power UCN sources (preferentially at the FRMII). We expect to
store about $10^{8}$ neutrons per measuring cycle in the volume of
700 l. Using the time spectrum of the decay protons a measuring
accuracy of 1 second can be achieved per cycle (about 40min)
\cite{brocker}. Thus, the goal of $\delta t_{stat}\sim 0.1s$ can be
achieved within 3-4 days. Much effort has been devoted to allow
enough flexibility to study systematics. Spin flip neutrons (if
created at all) can be detected with an efficiency of about 60\%,
stored neutrons with almost 100\% and decay protons with 70\%. The
momentum spectrum of the stored neutrons can be altered prior to
each measuring cycle as well as the spring tension (storing
B-field). We expect to operate this instruments in 2-3 years from
now.

\bigskip

\section{The Two-Body Decay of the Neutron $(n\rightarrow He^{-})$}

Although neutron decay is almost uniquely passing as three body decay ($%
n\rightarrow pe^{-}\overline{\nu })$ a small but important phase
space exists for the charged particle to form a hydrogen atom in
which case the kinematics of a two-body decay has to be considered.
This two-body kinematics bears a number of very interesting
features. If we consider massless neutrinos only, as well as
left-handed currents (standard model) the population of the
hyperfine states of the hydrogen atom is well predictable. In
addition, the direction of flight of the hydrogen atoms defines the
quantization axis for all angular momenta involved (when using
unpolarized neutrons). As the weak interaction is very short ranged,
all hydrogen atoms will be formed in a relative S-state (nS) with
the population of the various n-values to be governed again by phase
space and kinematics. The hyperfine state of the hydrogen atom is
directly related to the relative spin orientation of proton and
electron and thus to their helicities. If the decay happens inside a
small magnetic field, the helicities of all particles is conserved.
Thus, an analysis of the different possible hyperfine states gives
information about the validity of the assumptions made above.

Table 1 denotes the population of the different hyperfine states
\cite{Nemenov_80}. They are ordered in terms of helicities and will
mix, if no
magnetic field were present. As a convention, the H moves to the right, the $%
\bar{\nu}$ to the left. \textit{Fe} and \textit{GT} mean Fermi and
Gamov-Teller transition, respectively. $W_i$ are the populations
according to pure V-A interaction (\cite{Nemenov_80}\cite{Yer}), $F$
the total spin (with hyperfine interaction) and $m_F$ the $F$
projection, $\vert m_S m_I\rangle$ the Paschen- Back state, where
$m_S$ and $m_I$ denote the $e^-$ and $p$ spin quantum numbers (+
means +1/2, i. e., spin points to the left in the magnetic
quantization field direction).

In order to see the sensitivity to an admixture of other interactions (like
scalar or tensor) table two shows the modified population in case of such
small admixtures \cite{Nem2}.

\begin{table}
\label{tab1}

\begin{center}
\begin{tabular}{c|c|c|c|c|c|c|c|c|c}
config. i & $\bar{\nu}$ & n & p & $e^{-}$ & transition & $W_{i}$(\%) & $F$ &
$m_{F}$ & $|m_{S}m_{I}\rangle $ \\ \hline
1 & $\leftarrow $ & $\leftarrow $ & $\leftarrow $ & $\rightarrow $ & Fe & $%
44.174\pm .017$ & 0,1 & 0 & $|+-\rangle $ \\
2 & $\leftarrow $ & $\leftarrow $ & $\rightarrow $ & $\leftarrow $ & GT & $%
55.211\pm .013$ & 0,1 & 0 & $|-+\rangle $ \\
3 & $\leftarrow $ & $\rightarrow $ & $\rightarrow $ & $\rightarrow $ & Fe & $%
.615\pm .003$ & 1 & 1 & $|++\rangle $ \\
4 & $\rightarrow $ & $\leftarrow $ & $\leftarrow $ & $\leftarrow $ & Fe & 0.
& 1 & -1 & $|--\rangle $ \\
2' & $\rightarrow $ & $\rightarrow $ & $\rightarrow $ & $\leftarrow $ & Fe &
0. & 0,1 & 0 & $|-+\rangle $ \\
1' & $\rightarrow $ & $\rightarrow $ & $\leftarrow $ & $\rightarrow $ & GT &
0. & 0,1 & 0 & $|+-\rangle $%
\end{tabular}
\end{center}
\caption {Spin projections i in the neutron bound $\beta$ decay. As
a convention, the H moves to the right, the $\bar{\nu}$ to the left.
$W_i$ are the populations according to pure V-A interaction, $F$ the
total spin (with hyperfine interaction) and $m_F$ the $F$
projection, $\ket{m_S m_I}$ the Paschen-Back state, where $m_S$ and
$m_I$ denote the $e^-$ and $p$ spin quantum numbers.}
\end{table}

The absence of a direct population of configuration 4 is striking. One can
show, that this helicity configuration can only be populated if either the
neutrino mass is finite or if a right handed admixture to the weak
interaction exists. The latter can be formulated quantitatively by the
following expressions. If we only assume the existence of a right handed $%
W_{R}$ with mass $M_{R}$ then the corresponding coupling constant $g_{R\text{
}}$scales with $M_{R}^{2}$, the relative probability with $%
M_{L}^{2}/M_{R}^{2}$ multiplied with a factor $(1+\lambda
)^{2}/(2(1+3\lambda ^{2}))$ $\ll 1$ from spin gymnastics \cite{Kaiser05}.

\begin{table}
\label{tab2}

\begin{center}

\begin{tabular}{c|c|c|c}
config. i & $g_S = 0, g_T =0$ & $g_S = 0.1, g_T =0$ & $g_S = 0, g_T =0.02$
\\ \hline
1 & 44.174 & 46.479 & 43.440 \\
2 & 55.211 & 53.288 & 55.789 \\
3 & .615 & .233 & .771 \\
4 & 0. & 0. & 0. \
\end{tabular}
\caption{$W_i$(\%) for various $g_S$ and $g_T$.}
\end{center}
\end{table}

\bigskip

We may also consider left right symmetric models \cite{Holstein_1977qn}
where two mass eigenstates $W_{1\text{ }}$and $W_{2}$\ mix to form $W_{L%
\text{ }}$and $W_{R\text{ }}$with a mixing angle $\xi $ and mass ratio $\eta
=(W_{1\text{ }}/W_{2\text{ }})^{2}$. Then the contribution to configuration
4 will be

\begin{center}
$W_{4}=2(\eta -\xi +\lambda (\eta +\xi ))^{2}$
\end{center}

At any rate, $W_{4}$ is expected to be small, $W_{4}<10^{-7}.$ Thus a very
sensitive zero measurement is required where we have to also consider
feeding of this state by atomic cascading from higher lying S-states, though
populated with small probability.

\begin{figure}[ht]\begin{center}
\includegraphics[
height=4.2047in,
width=5.2572in]
{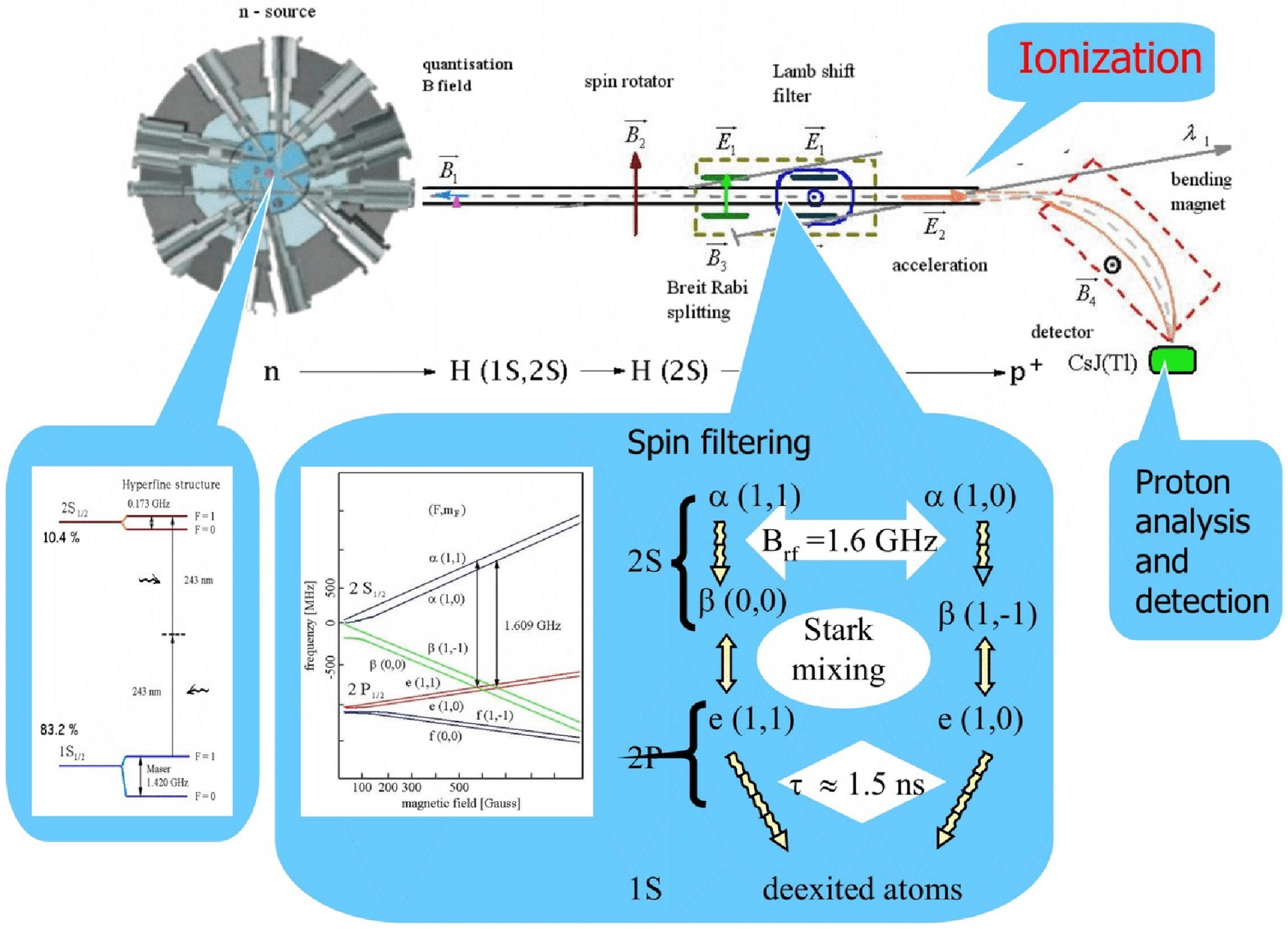}%
\end{center}\end{figure}%

The experimental method \cite{Dollinger}requires a strong neutron
source, a large decay volume and background free analysis of the
emerging hydrogen atoms. A possible setup is sketched in fig. 6
which depicts a typical research reactor with a tangential but
through-going beam tube acting as decay volume. Hydrogen atoms are
thus extracted without direct sight to the reactor core strongly
reducing background from emerging neutrons. After a few meters
flight path (within a small axial magnetic field) the hydrogen atoms
pass through a lamb-shift polarimeter, where one can select one
particular hyperfine state. The surviving hydrogen atoms are
detected by resonance ionization and subsequent detection of the
proton. The large Doppler-shift owing to the kinetic energy of the
hydrogen atoms of 323 eV is used to discriminate against rest gas
atoms.

Owing to presently existing laser powers in the UV range, only
hydrogen atoms produced in the n=2 meta-stable state can be used,
constituting about 10\% of all 2-body decays. The splitting of these
states in the analyzing field of the lamb shift filter is shown in
figure 6 (inlet) Using a high-power laser and a cavity we can induce
transitions between $\alpha $ and $\beta $-states by means of an rf
field of 1.609 GHz. Using an electric field of about 4.3 V/cm for
Stark-mixing we can control the quenching of the $\beta $-states to
the short living 2P states (e-states) and thus depopulate
selectively particular $\alpha $ and $\beta -$states
\cite{Haeberli_67}. Reversing the sign of the magnetic field in the
decay region, the association of the hyperfine states to $\alpha $
and $\beta $-states can be reversed allowing access to all possible
helicity configurations.

As there is no external trigger, the passage of a fast moving
hydrogen atom is detected via its two step ionization. In a first
step a Doppler-detuned laser induces the transition $2S\rightarrow
3P$ and a subsequent broad band light source ionizes the atom. The
resulting proton will be accelerated in an electric field and the
momentum analyzed in a spectrometer magnet equipped with a proton
detector.

We can estimate measuring accuracies assuming a typical neutron flux
as available at the FRMII in Munich of $2\cdot 10^{14} \rm
cm^{-2}s^{-1}$ \cite{Gau}, \cite{Alt}. This results in rate of
detectable hydrogen atoms of 3$ \rm s^{-1}$ of which 10\% are usable
(2S). Assuming high efficiencies $\varepsilon $
(thus high laser power) we can improve the current limits of $g_{S}$ and $%
g_{T}$ by a by a factor $2.5/$day$\times \varepsilon $. Present limits are $%
g_{S}$ $<6\cdot 10^{-2}$ \cite{Ade} and $g_{T}$ $<0.125$ \cite{Mum}. For the
forbidden decay, reachable limits are of the order of $10^{-6}$/yr.

\end{document}